\documentclass[preprint,pra,onecolumn]{revtex4-1}
\usepackage{amsmath}
\usepackage{amssymb}
\usepackage{array}
\usepackage{mathrsfs}
\usepackage{tabularx}
\usepackage{graphicx}
\usepackage{dcolumn}
\usepackage{bm}
\usepackage{graphicx}
\usepackage{color}
\usepackage{diagbox}

\newcommand{\tabincell}[2]{\begin{tabular}{@{}#1@{}}#2\end{tabular}}
\usepackage{multirow}
\usepackage{float}
\usepackage{appendix}
\usepackage{booktabs}
\usepackage{multirow}
\usepackage{epstopdf}

\begin{document}
	\title{Both qubits of the singlet state can be steered simultaneously by multiple independent observers via sequential measurement}
	\author{Kun Liu$^{1,2,3}$}
\author{Tongjun Liu$^{1,2,3}$}
\author{Wei Fang$^{1,2,3}$}
\author{Jian Li$^{1,2,3}$}\email{jianli@njupt.edu.cn}
\author{Qin Wang$^{1,2,3}$}\email{qinw@njupt.edu.cn}
	\affiliation{$^{1}$Institute of quantum information and technology, Nanjing University of Posts and Telecommunications, Nanjing 210003, China.}
	\affiliation{$^{2}$``Broadband Wireless Communication and Sensor Network Technology" Key Lab of Ministry of Education, NJUPT, Nanjing 210003, China.}
	\affiliation{$^{3}$``Telecommunication and Networks" National Engineering Research Center, NJUPT, Nanjing 210003, China.}
	\date{\today}
	\begin{abstract}
		Quantum correlation is a fundamental property which distinguishes quantum systems from classical ones, and it is also a fragile resource under projective measurement. Recently, it has been shown that a subsystem in entangled pairs can share nonlocality with multiple observers in sequence. Here we present a new steering scenario where both subsystems are accessible by multiple observers. And it is found that the two qubits in singlet state can be simultaneously steered by two sequential observers, respectively.
		
		PACS number(s): 03.67.Dd, 03.67.Hk,42.65.Lm

	\end{abstract}
	
	\maketitle	
	\section{Introduction}
	The correlation between the entangled distant quantum systems can be beyond any classical systems and plays an important role in both fundamental quantum physics and quantum information science. Quantum correlation is also a powerful resource in quantum information processes, it is widely applied in protocols such as teleportation, secure key distribution, quantum computation and etc. Generally, quantum correlations can be descripted as a hierarchy with three inequivalent concepts, quantum entanglement \cite{101}, Einstein-Podolsky-Rosen (EPR) steering \cite{102}, and Bell nonlocality \cite{103}. For bipartite systems, the weakest quantum correlation, entanglement, means that the state density cann't be decomposed into any separable one. Bell nonlocality, the strongest form, allows the violation of Bell inequality based on the joint probability distribution of local measurements by two spatially separated players. As the intermediate form, EPR steering, which was first introduced by Schr\"{o}dinger, and then formally defined by Wiseman \emph{et al.} \cite{10,17}, is verified if one of the two spatially separated parties can steer, via local measurements, the set of conditional quantum state of the subsystem on the other player's side.

For the certification of these correlations \cite{104}, the violation of Bell inequalities can verify nonlocality without any assumption of the measurement devices on both sides, which is considered as a device-independent scenario \cite{121}. Quantum entanglement can be certified via quantum state tomography with fully characterized measurement devices on both side, which is a device-dependent scenario. And for EPR steering, the measurement devices on one side need to be fully characterized, which is also called one-side device-independent scenario, or a semi-device-independent one \cite{3}.

Usually, quantum correlation is investigated for the $n$ particles in entanglement, each of which is accessible by one and only one space-separated observer for projective measurements \cite{123}. Some studies use different degree of freedom of parties. In Ref. \cite{105}, the steerability of three entangled qubits is demonstrated with two photons with polarization and spatial degree of freedom. However, it is interesting to consider whether multiple observers sequentially measuring the same subsystem can share different quantum correlation. Generally, for a sharp measurement, a system would collapse into one of the eigenstates of the projector, where the system would be separable from other systems. However, unsharp, or weak measurement can preserve some entanglement in the post-measurement state. Weak measurement refers to the measurement with intermediate coupling strength between the system and the probe. In contrast to strong projective measurement, weak measurement can be less destructive and also remain some original features of measured system. It has been proved to be a good method for signal amplification \cite{106}, state tomography and solving quantum paradoxes. For a pair of entangled particles, it is possible to reserve some quantum correlation after measurement at intermediate strength \cite{107}. Silva \emph{et al.} \cite{108} showed that multiple observers can share nonlocality on a pair of entangled qubits by using optimal weak measurement in a Bell scenario. One qubit is accessible by a single observer, Alice, while the other one is accessible by several observers, Bobs, sequentially. Without any communication between Bobs, the CHSH-Bell inequality between Alice and two Bobs, respectively, can be violated simultaneously. The violation of twice CHSH-Bell inequality is demonstrated \cite{14} using a pair of polarization-entangled photons. Brown and Colbeck \cite{109} shown that arbitrary number of Bobs can share the CHSH nonlocality with a single Alice with a single maximally entangled qubit pair, if, for each Bob, one measurement is sharp while the other is optimal weak. It brings a further understand of the fundamental limit of entanglement applications in device-independent scenarios. Foletto \emph{et al.} restudied the correlation between Alice and individual Bobs \cite{110}. that a system can still sustain entanglement and violate CHSH inequality if arbitrary number of observers (Bobs) implement sequential unsharp measurements on one subsystem and the other observer, Alice, performs implement sharp measurement (Alice). The measurements on the subsystem are based on the history of previously performed measurements and observed outcomes, so when appropriate measurement strength is choosed, the path of a tree-like structure (the protocol) will extend infinitely.

Shneoy\emph{et al.} \cite{111} theoretically investigated the EPR steering with multi-observer. it is found that with the increase of dimension in system, if measurement strength $\eta_{i}$ all have appropriate values, the number of that Bobs can steer Alice is found to be $N_{Bob}\sim d/\log d$. For the case of qubits $d=2$, each Bob knows his respective position in the sequence, as he performs a measurement with a well chosen strength $\eta_{i}$, the maximal sequence of Bob that can steer Alice is Five. And Choi \emph{et al.} \cite{12} experimentally demonstrated this process by accomplishing a sequential steering for three Bobs using photonic system.

To date, most discussions of quantum correlation for multiple observers with single entangled qubit pair focus on that one qubit is accessible by several Bobs sequentially, while the other one is accessible by one Alice. In this letter, we exploit the EPR steerability of a maximally entangled qubit pair, where each qubit is accessible by multiple observers sequentially. It is worth mentioning a similar scene in revealing of hidden nonlocality \cite{112,113}, where both qubits are locally filtered before the final sharp measurements. The local filter on each qubit can be considered as a special positive operator-valued measurement (POVM), where only one output is delivered to the successive observer, while the weak measurements above are one type of POVM with two outputs, both of which is accessible for all successive observers. The EPR steering scenario and linear steering inequality are presented as the basic tool in section II. A detailed steering protocol for multiple observers using weak measurement is given in section III. The violation of steering inequality for different Alice-Bob pairs is discussed followed by the conclusion in last section.
\section{EPR steering and steering inequality}\label{sec2}
	\subsection{EPR Steering scenario}\label{primary}
	Considering a pair quantum subsystems, $A$ and $B$, is shared by two spatially-separated observers, Alice and Bob, where each subsystem is accessible by one and only one observer. Generally said, steering describes the ability that one observer, Alice (Bob), via the measurement on $A$ ($B$), can nonlocally steer the other's subsystem $B$ ($A$).
	\begin{figure}[h!]
		\begin{center}
			\includegraphics[width=0.4\linewidth]{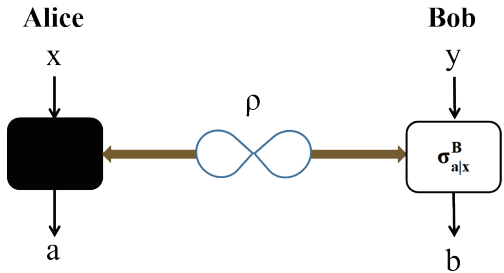}
		\end{center}
		\caption{General scenario in quantum steering. Alice and Bob are two spatially separated observers.
	The measurement device of Alice can be represented by a black box, with the classical input $x$
	(choice of measurements) and output $a$. Bob performs tomography and records his own
	measurement choice $y$ and the result $b$.}
		\label{fig1}
	\end{figure}

To be specific, quantum steering can be described as a scenario with two spatially-separated players, Alice and Bob,  sharing a bipartite quantum state $\rho_{AB}$. Alice implements a measurement with the inputs $x$ on her subsystem $A$ and outcomes $a$. Both $x$ and $a$ are classical variables. Bob can fully characterize the state in his subsystem with trusted devices. Conditioned on Alice's declare of $(x,a)$ with probability $p(a|x)$, Bob would found his subsytem $B$ collapsed into the state $\tilde{\sigma}_{\mathrm{a|x}}^{B}$ (unnormalized) \cite{124}:
\begin{align}\label{sig}
\tilde{\sigma}_{\mathrm{a} \mid \mathrm{x}}^{B}=t r_{A}\left(\mathrm{M}_{a \mid \mathrm{x}} \otimes \mathrm{I}_{\mathrm{B}} \cdot \rho_{A B}\right),
\end{align}
The normalized state is $\sigma_{\mathrm{a|x}}^{B}=\frac{\tilde{\sigma}_{\mathrm{a} \mathrm{x}}^{B}}{p}$, where $p=\operatorname{Tr}\left(\tilde{\sigma}_{\mathrm{a|x}}^{B}\right)$. The set of Bob's conditioned states for all measurement choice and results of Alice forms an assemblage. The assemblage obey a local hidden state (LHS) model, if each state from the assemblage can be written as,
\begin{align}\label{sigmaB}
\sigma_{\mathrm{a} | x}^{\mathrm{B}}=\int f(\lambda) P_{\mathrm{A}}(a | \mathrm{x}, \lambda) \sigma_{\lambda} \mathrm{d} \lambda
\end{align}
where, $\lambda$ is a random variable with the distribution $f(\lambda )$ as a strategy of Alice. Then Alice cannot steer Bob's subsystem as the state conditioned on $(x,a)$ can be chosen from the pre-existing ensemble $\{\sigma_{\lambda}\}$ with the distribution $f(\lambda)$ by Alice's declaration probability $\mathrm{P}_{\mathrm{A}}(a \mid x, \lambda)$  \cite{111}. Otherwise, if the assemblage cannot adpot any LHS model, Bob can be convinced that his subsystem can be steered by Alice remotely.
And steerability could be one-directional, where only one player (Alice) can steer the subsystem accessible by the other one (Bob) \cite{114}. Here, we focus on the bidirectional steering scenario and show that due to the existence of sequential and weak measurement, both Alices and Bobs can steer the remote subsystem.

	\subsection{Steering Inequality for qubits}\label{Inequality}
The EPR steerability can be verified by the violation of linear steering inequalities \cite{115}, just like the Bell inequality for Bell nonlocality. In the scenario, Alice declare her result $a_{k}\in\{-1,+1\}$ for each choice of measurement $k\in \{1,2,\ldots,n\}$, and Bob makes the corresponding dichotomic measurement $\hat{B}_{k}$ on his subsystem. The inequality reads:
\begin{align}\label{Sn}
\mathrm{Sn}=\frac{1}{\mathrm{n}} \sum_{\mathrm{k}=1}^{n}\langle a_{k} \hat{B}_{k}\rangle \leq \mathrm{Cn}
\end{align}
$C_n$ is the upper bound of inequality allowed by the LHS model,
\begin{align}\label{Cn}
\mathrm{Cn}=\max _{\{\mathrm{a_k}\}}\left\{\lambda_{\max} \left(\frac{1}{\mathrm{n}} \sum_{k=1}^{n} a_{k} B_{k}\right)\right\}
\end{align}
where $\lambda_{\max }(\mathrm{G})$ is the maximum eigenvalue of operator $G$.
To verify the steerability, both players can perform dichotomic quantum measurement, $\{\hat{A}_{k}\}$ and $\{\hat{B}_{k}\}$ on the qubit $A$ and $B$ respectively, can calculate the expected value of steerability parameter $S_{n}$ from the statistics. If the inequality is violated (satisfied $S_n$ exceed $C_n$), it means that Alice can steer Bob and Bob can also steer Alice, vice versa. And the LHS bound $C_{n}$ changes with the number of measurements allowed, $\mathrm{C} 2=1 / \sqrt{2} \approx 0.7071$, $\mathrm{C} 3=\mathrm{C} 4=1 / \sqrt{3} \approx 0.5773$, $\mathrm{C} 6 \approx 0.5393$, $\mathrm{C} 10 \approx 0.5236$. The lower the bound is , the more robust to noise of the steering phenomenon.
	\section{Multi-observer steering via sequential measurement }\label{sec3}
\begin{figure}[hptb]
		\begin{center}
			\includegraphics[scale=0.3]{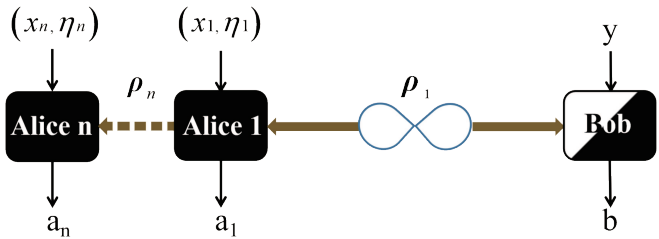}
		\end{center}
		\caption{The scenario of $1$ vs $n$ nonlocality (steering) verification. State $\rho$ is a shared state, one for all $n$ Alices and the other only for Bob. $Alice_1$$\sim$$Alice_{n-1}$ implement weak measurement with strength $\eta_1$$\sim$$\eta_{n-1}$, $Alice_n$ and Bob implement strong measurement. Bob, half black and half white, represents two modes in different conditions. In Bell-nonlocality scenario, black box denotes no preassumption on the measurement, which is device-dependent. However, in steering scenario, white box allows Bob to perform trusted quantum measurement or state tomography, which is device-independent.}
		\label{fig2}
	\end{figure}

	In most study on multi-observer nonlocality with single entangled pair, one subsystem is measured sequentially by different observers, while the other one is accessible by only one observer. Here we call it the $1$ vs $n$ nonlocality. The model is shown in figure \ref{fig2}. For multiple steering protocol, it is the one directional steering on the same subsystem. In Ref. \cite{111}, it is shown that for optimal weak measurements of all Alices, the number of that Alices can steer Bob is found to be $N_{Alice}\sim d/\log d$. For qubit system, $d=2$, the maximal number of Alices is $5$.
	
\begin{figure}[hptb]
		\begin{center}
			\includegraphics[scale=0.3]{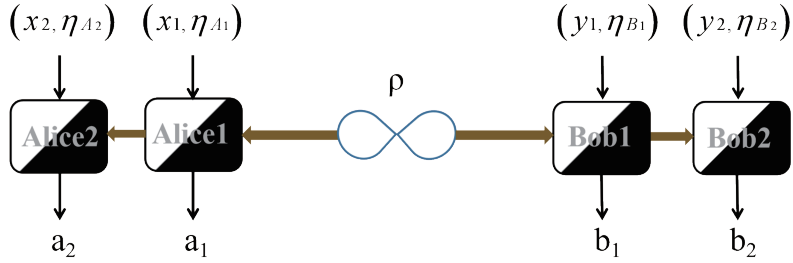}
		\end{center}
		\caption{A new type of quantum network in this paper. Two subsystem share state $\rho$, $Alice_1$ and $Bob_1$ implement weak measurement and the parameter strengths are $\eta_{A_1}$ and $\eta_{B_1}$. $Alice_2$ and $Bob_2$ implement strong measurement. The box with black and white shows the observer can steer other and also can be steered by others.}
		\label{fig3}
	\end{figure}

	In our protocol, the maximally entangled two subsystems $A$ and $B$, are sequentially accessible by two observers, ($Alice_{1}, Alice_{2}$) and ($Bob_{1}, Bob_{2}$), respectively. On the one side with subsystem $A$, $Alice_1$ randomly selects its input $k\in\{1,2,\ldots,n\}$, performs a weak measurement $A_{1}(k,\eta_{A})$ with the sharpness $\eta_A\in[0,1]$), keeps the outcome $a_{1}$ and delivers the post-measurement subsystem $A$ to $Alice_{2}$. Here, $Alice_2$ have no knowledge of the measurement choice or the outcome of $Alice_{1}$. Then with the randomly choice of $j\in\{1,2,\ldots,n\}$, $Alice_{2}$ performs her measurement $A_{2}(j)$. As there is no further observer of $A$, the measurements by $Alice_{2}$ always sharp ones. The process is similar on the other side with subsystem $B$, where $Bob_{1}$ performs weak measurements and $Bob_{2}$ performs the sharp ones (see Fig.\ref{fig3}).
	
All the dichotomic measurements implements by four the players can be represented by POVMs with two Kraus operators.
Being a frequently-used operator in this task, a set of Krause operators can be defined as:
\begin{align}\label{Kn}
\mathrm{K}_{\pm \mid k}=\frac{1}{\sqrt{2}}\left[\sqrt{1 \pm \eta} \cdot\left(\frac{\mathrm{I}+\mathrm{m}_{k} \cdot \boldsymbol{\sigma}}{2}\right)+\sqrt{1 \mp \eta} \cdot\left(\frac{\mathrm{I}-\mathrm{m}_{k} \cdot \boldsymbol{\sigma}}{2}\right)\right]
\end{align}
Here, $k$ is $k$-th measurement, $\pm$ are the outcomes of dichotomic measurement, and $\eta$ is the strength of the measurement strength for $Alice$s or $Bob$s. The observable of POVM can be written as \cite{111}:
\begin{align}\label{Ai}
\hat{\mathrm{E}}_{k}=\mathrm{M}_{+|k}-\mathrm{M}_{-| k}=\eta \mathrm{m}_{k} \cdot \sigma
\end{align}
where
\begin{align}\label{M}
M_{\pm \mid k}=K_{\pm \mid k}^{\dagger} \cdot K_{\pm \mid k}
\end{align}

Here, $\sigma$ represents three Pauli matrixes $(\sigma_{x}, \sigma_{y}, \sigma_{z},)$, and $m_{k}$ denotes a unit vector on Bloch sphere. In other word, the set all observables can be represent by $n$ vectors $\{m_{k}\}_{k\in\{1,2,\ldots,n\}}$ and the measurement strength $\eta$, for each player. The measurement is sharp for $\eta=1$.  And for $\eta=0$, both Kraus operators are proportional to the identity, where the system is unchanged and no information can be inferred from the measurement.

To verify the steerability, the choice of measurement set should be optimised. Saunders \emph{et al.} \cite{115} have shown that, for the singlet state $|\Psi^{-}\rangle=\frac{1}{\sqrt{2}}(|01\rangle-|10\rangle$, the optimal measurements can be represented by the vectors $\{m_{k},-m_{k}\}$ related to the Platonic-solid depending on $n$, the number of measurements allowed for each player. Measurement settings correspond to different solids in Bloch space, octahedron ($n=3$), cube ($n=4$), icosahedron ($n=6$) and dodecahedron ($n=10$), except the case of $n=2$, where the four vectors form a square in a plane. Antipodal pairs of face centres or vertices of a Platonic solid define measurement vectors $\{m_{k}\}$. As for $n=2,3,4$, the vectors are from the origin of coordinate to the face centres, while for $n=6, 10$, to the vertices of solid \cite{122}.
	
	\begin{figure}[hptb]
		\begin{center}
			\includegraphics[scale=0.2]{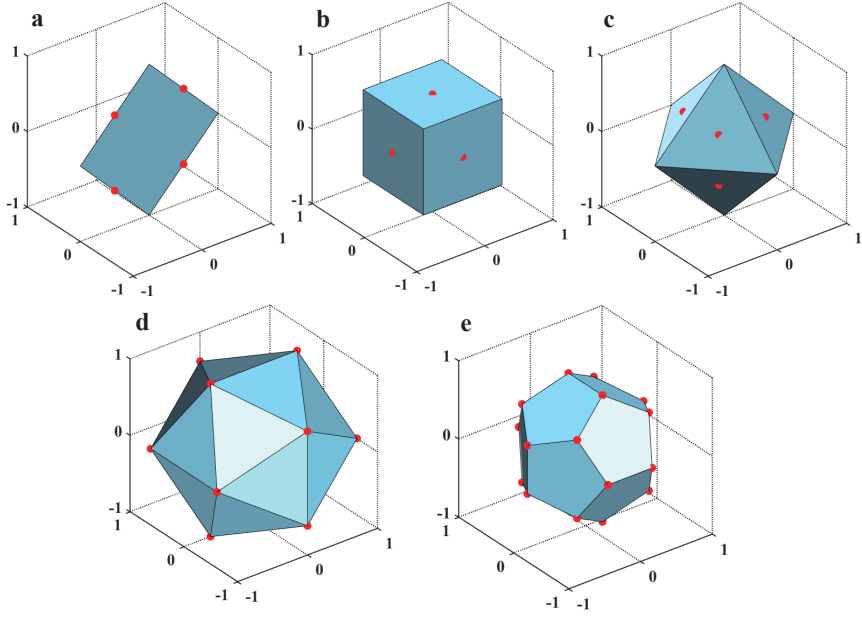}
		\end{center}
		\caption{The bullet symbols show the orientations of pure states in optimal cheating ensembles for two-qubit states\cite{115}. From a to e, we can get different pair of coordinates from $n=2,3,4,6, 10$.}
		\label{fig4}
	\end{figure}
	
Let's consider the steering scenario. The two qubits in state $\rho^{i'j'}$ is delivered to a pair of observers, $A$ to $Alice_{i}$ and $B$ to $Bob_{j}$, respectively. $Alice_{i}$ performs the measurement according the random input $k$ with the Kraus operators $\{K_{+|k}^{\eta_{A}},K_{-|k}^{\eta_{A}} \}$ and gets an output $a\in\{+,-\}$. And $Bob_{j}$ performs the measurement according the random input $l$ with the Kraus operators $\{K_{+|l}^{\eta_{B}},K_{-|l}^{\eta_{B}} \}$ and gets an output $b\in\{+,-\}$. The (unnormalized) conditional post-measurement two-qubit state would be expressed as,
 \begin{align}\label{unrho}
 \rho_{a|k,b|l}^{ij}=K_{a|k}^{\eta_{A}} \otimes K_{b|l}^{\eta_{B}} \rho^{i'j'}  K_{a|k}^{\eta_{A}\dagger} \otimes K_{b|l}^{\eta_{B}\dagger}.
\end{align}
with the conditional probability $p(a,b|k,l) = \mathrm{Tr} \rho_{a|k,b|l}^{ij}$. So the steerability parameter for $Alice_{i}$ and $Bob_{j}$ is written as,
 \begin{align}\label{Sij}
\hat{S}^{i j}_{n}=\frac{1}{n} \sum_{k=1}^{n} \sum_{a, b\in\{+,-\}}(-1)^{a+b}  p\left(a, b|k,k\right).
\end{align}
And the average post-measurement state after $Alice_{i}$'s and $Bob_{j}$'s measurements is
 \begin{align}\label{rho}
\rho^{ij} = \frac{1}{n^{2}}\sum_{k,l=1}^{n}\sum_{a,b\in\{+,-\}} \rho_{a|k,b|l}^{ij}.
\end{align}
which can be delivered to subsequential observers for further investigation.
In our multi-observer steering scenario, the initial state is in singlet one, $\rho^{00}=|\Psi^{-}\rangle\langle\Psi^{-}|$. If $S^{ij}_{n} > C_{n}$, the bidirectional steerability can be declared between
 $Alice_{i}$ and $Bob_{j}$ from state $\rho^{i'j'}$, where $i'=i-1$ and $j'=j-1$, and delivering the post-measurement state $\rho^{ij}$. To investigate the steerability between $Alice_{i}$ and an observer after $Bob_{j}$, $Alice_{i}$ has to postpone her measurement. Equivalently, she can perform the weak measurements with the sharpness $\eta_{a}=0$ and keep the post-measurement subsystem $A$ for further measurements.
\begin{center}
\begin{table}[htbp]
  \caption{Measurement directions for different settings. (a=(1+$\sqrt{5}$) / 2)}
  \setlength{\tabcolsep}{5mm}
   {\begin{tabular}{cc}

\hline \hline
Measurement setting n & (X,Y,Z) \\
\specialrule{0.05em}{2pt}{2pt}
2& (1,0,1),(1,0,-1)\\
\specialrule{0em}{2pt}{2pt}
3& (1,0,0),(0,1,0),(0,0,1)\\
\specialrule{0em}{2pt}{2pt}
4& (1,1,1),(1,-1,-1),(1,1,-1),(1,-1,1)   \\
\specialrule{0em}{2pt}{2pt}
6& (0,1,a),(0,1,-a),(1,a,0),(1,-a,0),(a,0,1),(a,0,-1) \\
\specialrule{0em}{2pt}{2pt}
10& \tabincell{c}{(0,1/a,a),(0,1/a,-a), (1/a,a,0),(1/a,-a,0)\\(a,0,1/a),(a,0,-1/a),
(1,1,1),(1,-1,-1),(1,1,-1),(1,-1,1)}\\
\hline \hline

   \end{tabular}}
   \label{table1}
\end{table}
\end{center}

\section{Result and Conclusion}\label {4}
For our multi-observer steering scenario, each subsystem is accessible by two players respectively. So the steering should be investigated under four observer-pairs, $S^{11}$, $S^{12}$,$S^{21}$, and $S^{22}$. Let the two qubit initially in the singlet state, we numerically calculate the steerability parameters.

		\begin{figure}[hptb]
		\begin{center}
			\includegraphics[scale=0.8]{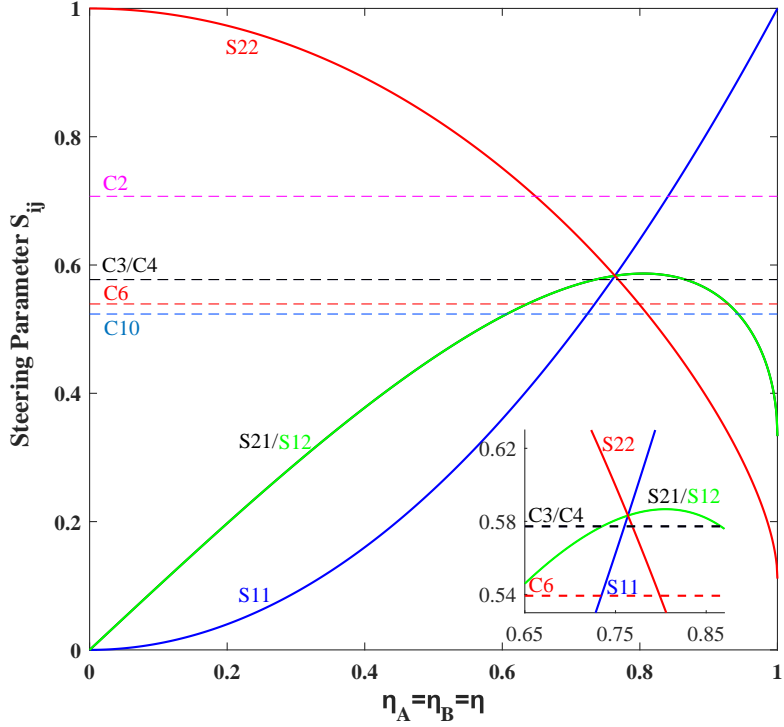}
		\end{center}
		\caption{Theoretical predictions of EPR steering correlation $S_n$ ($S_{ij}$) changing with the sharpness $\eta$, where we set $\eta_A=\eta_B=\eta$. Steering parameters $S_{ij}$ is function of $\eta$, $\eta\in(0,1)$ denotes the equal strength of measurement in Alice and Bob. The blue solid curve represents S11 ($Alice_1 \&Bob_1$) in theoretical predictions, because S12=S21 is still true, black (green) solid curve represents S21/S12 ($Alice_2\&Bob_1$/ $Alice_1\&Bob_2$), and the red one represents S22 ($Alice_2\&Bob_2$); The magenta dashed horizontal line represents steering bound C2, and the black one shows C3 (C4), red dashed line represents C6, blue dashed line represents C10.}
		\label{fig5}
	\end{figure}
\begin{figure}[hptb]
		\begin{center}
			\includegraphics[scale=0.8]{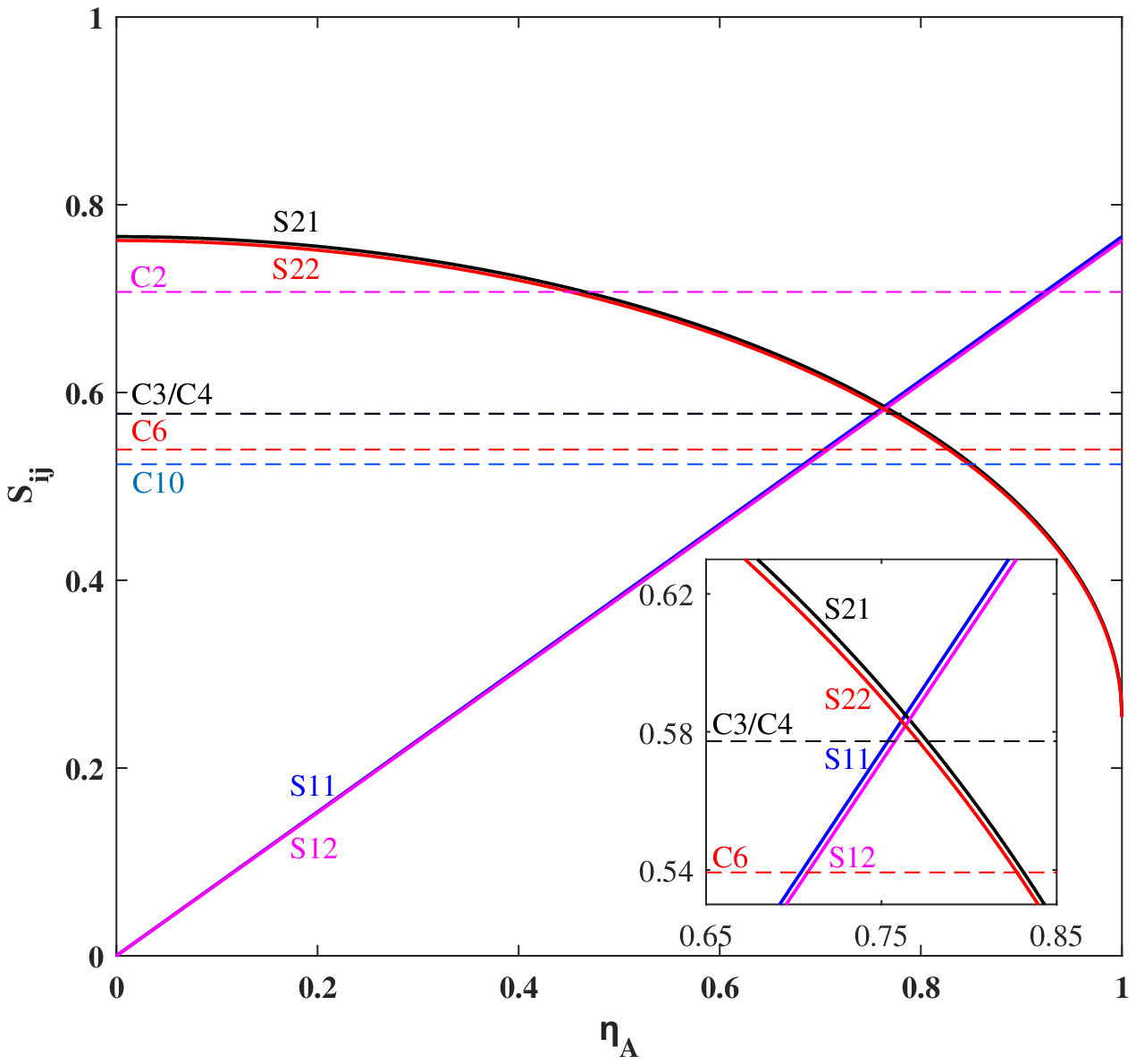}
		\end{center}
		\caption{The variation of $S_{ij}$ with $\eta_A$, conditional on fixed $\eta_B$. Value $S_{ij}$ is the function of variable $\eta_A$. Only one variable affects the violations of different measurement setting n. Herein, S11, blue solid line; S21, black solid line; S12, magenta solid line; S22, red solid line; C2, magenta dashed line; C3/C4 black dashed; C6, red dashed line; C10, blue dashed line.}
		\label{fig6}
	\end{figure}

Considering that both $Alice_{1}$ and $Bob_{1}$ perform independent measurement with the same sharpness, $\eta_{A}=\eta_{B}=\eta$. The theoretical predictions of EPR steering correlation $S_n$ ($S_{ij}$) and the violation of steering inequalities are showed in Fig. \ref{fig5}. The steerability parameters are calculated with a varying strength $\eta\in[0,1]$ using the measurement set based on the Platonic solid. It is interesting that the steerability parameters are independent of $n$, the size of the measurement set. And due to the symmetry between $Alice$s and $Bob$s, it is found that $S^{12}=S^{21}$. When both $Alice_{1}$ and $Bob_{1}$ perform the sharp measurement, only the observer-pair ($Alice_{1}, Bob_{1}$) can steer each other.   When both $Alice_{1}$ and $Bob_{1}$ perform the zero-sharpness measurement, only the observer-pair ($Alice_{2}, Bob_{2}$) can steer each other. Around $\eta\sim 0.76$, it is found that all the four observer-pairs can violate against the linear steering inequalities for $n=3,4,6,10$. That means, every player can steer the state of the remote subsystem. The steerability for the observer-pair ($Alice_{1}, Bob_{1}$) and ($Alice_{2}, Bob_{2}$) is a trade-off with $\eta$, while intermediate between them for ($Alice_{1}, Bob_{2}$) and ($Alice_{2}, Bob_{1}$).
In Fig. \ref{fig6}, we fixed the sharpness of $Bob_{1}$'s measurements $\eta_{B}=0.766$. It is found that the steering ability of two $Bob$s is very close, while being a trade-off between two $Alice$s with $Alice_{1}$'s  measurement strength $\eta_{A}$.
	
In conclusion, a new type of steering scenario with a single pair of entangled subsystem is presented in this article, each of which is accessible by multiple observers. We also showed that, by using weak measurements under the Platonic-solid configure, all the four independent observers, two sequentially measuring on the subsystems respectively, can steer the remote subsystem. Besides, when fixing $Bob_{1}$'s measurement strength as above, it would be more interesting to study the asymmetry of the steerability originating from the entangled state and other measurement sets. In such scenario, it's also worth studying how many observers can steer the remote subsystem if high-dimensional entanglement is shared.
Furthermore, recently, it is found that the sequential multi-party quantum random access code can witness the quantum channel \cite{20}, verify the unsharp measurement \cite{116,117}, and be generalized to characterise the correlations of quantum network in prepare-and-measurement scenarios \cite{118}. We believe that our steering scenario can also be implemented for quantum communication in similar network.

\begin{acknowledgments}
This work was supported by National Key Research and Development Program of China (2018YFA0306400, 2017YFA0304100), National Natural Science Foundation of China (12074194, 11774180, U19A2075), the Leading-edge technology Program of Jiangsu Natural Science Foundation (BK20192001).
\end{acknowledgments}

\bibliography{New_arxiv}

\end{document}